\documentclass[12pt]{article}
\usepackage{amssymb}
\usepackage{amsmath}
\usepackage{latexsym}
\usepackage{epsfig}
\usepackage{graphicx}
\usepackage{subfigure}
\usepackage{wasysym}
\usepackage{threeparttable}
\usepackage{natbib}
\usepackage{color}
\usepackage{epstopdf}
\usepackage{setspace}
\newcommand{\mbv}[1]{\mbox{\boldmath$#1$\unboldmath}}
\newcommand{\mbf}[1]{\mathbf{#1}}

\usepackage{hyperref}

\def\boxit#1{\vbox{\hrule\hbox{\vrule\kern6pt
          \vbox{\kern6pt#1\kern6pt}\kern6pt\vrule}\hrule}}

\def\bse{\begin{eqnarray*}}
\def\ese{\end{eqnarray*}}
\def\be{\begin{eqnarray}}
\def\ee{\end{eqnarray}}
\def\bq{\begin{equation}}
\def\eq{\end{equation}}
\def\bse{\begin{eqnarray*}}
\def\ese{\end{eqnarray*}}

\begin{document}
\thispagestyle{empty} \baselineskip=28pt

\begin{center}
{\LARGE{\bf Small Area Estimation via Multivariate Fay-Herriot Models with Latent Spatial Dependence}}
\end{center}

\baselineskip=12pt

\vskip 2mm
\begin{center}
Aaron T. Porter\footnote{(\baselineskip=10pt to whom correspondence should be addressed) Department of Statistics, University of Missouri, 146 Middlebush Hall, Columbia, MO 65211, porterat@missouri.edu},
Christopher K. Wikle\footnote{\baselineskip=10pt Department of Statistics, University of Missouri, 146 Middlebush Hall, Columbia, MO 65211},
Scott H. Holan$^2$
\end{center}
%
%
%
%
\vskip 4mm

\begin{center}
\large{{\bf Abstract}}
\end{center}

The Fay-Herriot model is a standard model for direct survey estimators in which the true quantity of interest, the superpopulation mean, is latent and its estimation is improved through the use of auxiliary covariates.  In the context of small area estimation, these estimates can be further improved by borrowing strength across spatial region or by considering multiple outcomes simultaneously.  We provide here two formulations to perform small area estimation with Fay-Herriot models that include both multivariate outcomes and latent spatial dependence. We consider two model formulations, one in which the outcome-by-space dependence structure is separable and one that accounts for the cross dependence through the use of a generalized multivariate conditional autoregressive (GMCAR) structure. The GMCAR model is shown in a state-level example to produce smaller mean square prediction errors, relative to  equivalent census variables, than the separable model and the state-of-the-art multivariate model with unstructured dependence between outcomes and no spatial dependence.  In addition, both the GMCAR and the separable models give smaller mean squared prediction error than the state-of-the-art model when conducting small area estimation on county level data from the American Community Survey.

\baselineskip=12pt

%
%
%

\baselineskip=12pt
\par\vfill\noindent
\par\medskip\noindent
\clearpage\pagebreak\newpage \pagenumbering{arabic}
\baselineskip=24pt
\section{Introduction}\label{sec:Intro}

It is often desirable to obtain estimates of population parameters in small spatial areas that are not sampled or are under-sampled in surveys. In such cases, small area estimation (SAE) is facilitated by the introduction of covariates that inform the population parameters and thus reduce estimation variance.  In its basic form, the Fay-Herriot (FH) model \citep{fay1979estimates} is a two-level hierarchical model in which the direct survey estimates are conditioned on a latent population parameter of interest (or survey ``superpopulation mean") and a survey variance, and the latent parameter is ``regressed'' on relevant covariates. Specifically, the FH model can be expressed as
\begin{eqnarray}
\label{eq:fhcore}
Y_{i}=\theta_{i}+\epsilon_i , \\
\label{eq:fhaux}
\theta_i =\mbf{x}_i' \mbv{\beta} +u_i,
\end{eqnarray}
where $Y_i$ denotes a design-unbiased estimate of $\theta_i$, the parameter of interest at location $i=1,\ldots,n$. In general, the error term $\epsilon_i$ in (\ref{eq:fhcore}) is assumed to be independent, normally distributed with mean zero and variance $\sigma_{i}^{2}$, with $\sigma_{i}^{2}$ representing the sampling variance at location $i$. This survey variance is often assumed known (provided by the official statistical agency) and the normality assumption may be relaxed \citep[e.g.,][]{you2002small}. In the second model stage (\ref{eq:fhaux}), auxiliary information, through the covariates $\mbf{x}_i$, can be utilized to provide reduction of the total mean squared error (MSE). The term $u_i$ denotes a spatially indexed random effect and, in the traditional model, is assumed to follow an independent normal distribution with unknown variance $\sigma_{u}^{2}$. Importantly, by modeling $\theta_i$ through equation (\ref{eq:fhaux}), the model effectively uses the auxiliary covariate information at location $i$ 
to draw strength across locations.  Such synthetic estimators make use of this auxiliary information in order to reduce the variance of the estimate of the small area parameter of interest, $\theta_i$, but at the expense of a potential introduction of bias \citep{rao2003small} -- the usual variance/bias tradeoff that is common in statistical inference.

%


In addition to using covariates to facilitate SAE in the classic FH model, one can also borrow strength across the small areas themselves, making use of the spatial dependence to improve estimates as is commonly done in spatial statistics \citep[e.g., see][]{cressie93, cressiewikle2011}. That is, one considers spatial dependence in $\{u_i\}$, which correspond to geographical areas.
The use of spatial dependence structures in SAE is not new, with models utilizing conditional autoregressive (CAR) spatial structure that is prevalent in the disease mapping literature  \citep[e.g.,][]{leroux1999estimation, macnab2003hierarchical}, which builds upon the classic work of \cite{clayton1987empirical}, \cite{cressie1989spatial}, \cite{mollie1991empirical}, \cite{clayton1992bayesian}, among others. An overview of the theory and application of CAR models can be found in \cite{rue2005gaussian}.  Such models have also been used in the FH context, with spatial structures given by CAR, intrinsic CAR (ICAR) and simultaneous autoregressive (SAR) models \citep[e.g.,][]{cressie1990small, pratesi2008small, gomez2010bayesian, you2011hierarchical}.  The majority of spatial FH models in the literature that utilize the CAR or ICAR structure consider are univariate.


In principle, one can also improve SAE by making use of other survey outcomes that are related to the primary outcome of interest.  Indeed, such multivariate FH models have been developed to account for correlation between the survey estimates of several parameters \citep{fay1987application}. Typically, one can attempt to account for the multivariate dependence either through the sampling error term (i.e., $\{\epsilon_i\}$ in (\ref{eq:fhcore})) or through the latent error term (i.e., $\{u_i\}$ in (\ref{eq:fhaux})). Most of the multivariate models in the literature have been concerned with modeling within-location dependence in the sampling errors \citep[e.g.,][]{fay1987application,huang2004empirical, gonzalez2008analytic,fabrizi2011hierarchical}. In some cases, this sampling error dependence is assumed known {\it a priori} \citep[e.g.,][]{fay1987application,huang2004empirical, gonzalez2008analytic} but this can be estimated in a hierarchical Bayesian framework with priors (e.g., inverse Wishart (IW) priors) on the within-location sampling dependence. More generally, \cite{datta1998multivariate} and  \cite{kubokawa2011parametric}  consider the possibility of dependence at both the sampling and latent error levels.  However, these formulations tend to over-generalize the modeling of the dependence in the latent structure, so that explicit spatial dependence is difficult to estimate due to potential confounding with the dependence in the sampling errors.  Finally, \cite{torabi2011hierarchical} considers a univariate outcome that varies with space and time.  That is, the model includes a separable spatio-temporal dependence structure with a  univariate CAR model on the spatial locations along with temporal dependence that is then necessarily assumed to be the same for all spatial locations.

Given that spatial dependence and multivariate dependence have been shown to improve estimation in FH SAE implementations, it follows that allowing for multivariate spatially dependent error structures in this framework could borrow strength across responses and through residual spatial dependence, thus providing improved estimators in the sense of reduction in MSE.  Further, because the FH model utilizes data on an irregular lattice, a sensible way of constructing dependence within the latent random effects for the multivariate FH model would be to extend the CAR structure that is sometimes used in the univariate FH model to the multivariate case.

The first multivariate CAR model was due to \cite{mardia1988multi}, but this model is very general and can be difficult to implement in practice. Other useful models followed, including \cite{kim2001bivariate}, \cite{gelfand2003proper}, \cite{jin2005generalized}, and \cite{sain2011spatial}. We propose two ways of handling the extension of the CAR structure to the multivariate FH model.  In the first, we consider a simple separable covariance structure that uses a single CAR model and builds multivariate dependence though a Kronecker product with an unstructured within-area covariance matrix (that is assumed to have an inverse Wishart (IW) prior in a Bayesian implementation). In the second, we use a fully multivariate CAR structure.   We note that the model of \cite{sain2011spatial} is nonseparable. However, the restrictive parameterization of the variances in this model makes it undesirable for the current application. Instead, we consider the GMCAR model of \cite{jin2005generalized} to jointly model the multivariate spatial dependence within the latent random effects. Importantly, this GMCAR model contains many multivariate CAR models as special cases, giving it the flexibility to account for a wide variety of spatial correlation structures in a nonseparable model and it is practical to include within a hierarchical latent effect framework.  Thus, this model provides an important extension of the recent interest in CAR and ICAR priors and multivariate dependence in the FH framework by providing natural extensions to FH spatial structure in higher dimensions.  Utilizing the GMCAR and separable spatial priors, we demonstrate, by way of two data analyses, that multivariate spatial priors can be effectively used to reduce both the posterior variance and the MSE of FH model estimates as compared to the traditional multivariate FH methodology. 

We begin in Section~ \ref{sec:MCARFH} with a discussion of the proposed multivariate spatial FH methodology.  We provide two simulated examples in Section~ \ref{sec:sim}. Then, to compare to a situation in which we have census values, and consider these values as ``truth,'' we provide a state level analysis.  Although this is not strictly dealing with ``small areas,'' it does illustrate the methodology in a situation where we actually have a proxy for the truth and at a level of geography of great interest by practitioners.  This is followed in Section~ \ref{sec:data} by a county-level SAE application.  We conclude with a brief discussion in Section~ \ref{sec:disc}.

\section{Multivariate Spatial Fay-Herriot Models}\label{sec:MCARFH}

Before defining the multivariate spatial FH models, we introduce here notation for spatial CAR models as used in the spatial statistics literature. Specifically, the aforementioned CAR models are defined by a set of full conditional distributions, as developed in \cite{besag1974spatial} \citep[see also][for a recent overview]{rue2005gaussian}: 
\begin{eqnarray}
\label{eq:CAR}
u_i|u_j \sim Gau \left(\sum_{i \sim j} \rho w_{ij} u_j , \tau_i^2 \right),
\end{eqnarray}
where  $Gau(a,b)$ denotes a Gaussian random process that follows a normal distribution with mean $a$ and variance $b$, as is common in spatial statistics.  Further, the notation $i \sim j$ indicates that locations $i$ and $j$ are neighbors (typically, that they share a border), $\rho$ is a spatial dependency parameter, and $w_{ij}$ is an adjacency weight.  For our purposes, we set $w_{ij}=1/w_{i+}$ if locations $i$ and $j$ are neighbors and $0$ otherwise, where $w_{i+}$ is the sum of the neighbors of location $i$. When this parameterization is utilized, $\tau_i^2$ is defined as $\tau^2/w_{i+}$, which guarantees the self-consistency of $\mbf{u} \equiv (u_1,\ldots,u_n)'$, and ensures that a symmetric, nonnegative definite precision matrix results from the conditional specification. The spatial dependency parameter $\rho$ has support on (-1,1) in order to guarantee that the precision matrix is positive definite, resulting in a proper joint distribution of the form
\begin{eqnarray}
\mbf{u} \sim Gau \left(\mbf{0},\tau^{2} (\mbf{D}-\rho \mbf{W})^{-1} \right),
\end{eqnarray}
where we define ${\mbf W}$ to be a matrix with element $(i,j)$ taking the value $1$ if locations $i$ and $j$ are neighbors and $0$ otherwise, and ${\mbf D}$ to be a diagonal matrix with element $(i,i)$ being $w_{i+}$.

\subsection{Multivariate FH Models}

The univariate FH model is extended to the multivariate case by assuming $Y_{ij}$ is a direct survey estimator for the $j$th outcome ($j=1,\ldots,m$) at location $i=1,\ldots,n$. Then, a natural multivariate extension of (\ref{eq:fhcore}) and (\ref{eq:fhaux}) is given by
\begin{eqnarray}
\label{eq:mvfhcore}
Y_{ij}&=&\theta_{ij}+\epsilon_{ij}, \\
\label{eq:mvfhaux}
\theta_{ij} &=&\mbf{x}_{ij}' \mbv{\beta}_j +u_{ij},
\end{eqnarray}
where the parameters are defined as above, but are now indexed by the outcome, $j$.  Recall that the structure of the FH model in (\ref{eq:fhcore}) typically assumes a known sampling variance $\sigma_i^2$ at location $i$. Generalizing to the multivariate case, we could analogously assume a known sampling variance-covariance matrix  $\mbox{cov}({\mbv \epsilon}_i) \equiv {\mbf \Sigma}_i$ for location $i$, where ${\mbv \epsilon}_i = (\epsilon_{i1},\ldots,\epsilon_{im})'$ . It is typically assumed that there is independence in the sampling errors between locations, and so $\mbv{\epsilon}\equiv(\mbv{\epsilon}'_1,\ldots, \mbv{\epsilon}'_n)' \sim N(\mbf{0}, \mbf{\Sigma})$, where $\mbf{\Sigma} = \hbox{diag}(\mbv{\Sigma}_1,\ldots,\mbv{\Sigma}_n)$.  As mentioned in Section \ref{sec:Intro}, one could add dependence between locations in the sampling errors so that this covariance structure would not be block diagonal, but information surrounding the dependence in sampling errors between regions is typically not available. For this reason, we prefer to model such dependence in the latent error structure.

Analogous to the sampling errors, we define the latent error variance-covariance matrix $\mbox{cov}({\mbf u}_i) \equiv {\mbf \Sigma}_{u,i}$ for location $i$, where ${\mbf u}_i = (u_{i1},\ldots,u_{im})'$. In this case, we do not necessarily assume independence between spatial areas, and so we seek to model the joint $(m \times n)$-dimensional variance-covariance matrix of ${\mbf u} = ({\mbf u}'_1,\ldots,{\mbf u}'_n)'$, which we denote ${\mbf \Sigma}_u$.  For realistic sizes of $m$ and $n$, we will be unable to estimate an unstructured ${\mbf \Sigma}_u$ and thus, must find reasonable parameterizations.  \cite{datta1998multivariate} consider a useful, but fairly simple, structure given by ${\mbf \Sigma}_u = \mbf{I}_{n} \otimes \mbv{\Sigma}_{IW}$, where $\otimes$ denotes a Kronecker product and ${\mbf \Sigma}_{IW}$ is an $m$-dimensional unstructured covariance matrix between outcomes that is given an inverse-Wishart (IW) prior in their Bayesian implementation.  We consider this to be the ``state-of-the-art" latent dependence structure for multivariate FH models and refer to it as the ``IW-FH'' model.

Most multivariate FH models developed subsequent to \cite{datta1998multivariate} use a similar structure, accounting for within location correlation, but ignoring an explicit formulation for spatial structure that may be present in the latent spatially-referenced vector, $\mbf{u}$.  We propose to model the latent spatial structure in $\mbf{u}$ explicitly, in order to achieve a reduction in MSE for small area estimates. We describe two such structures below.

\subsection{Separable Multivariate Spatial FH Model}\label{sec:sepFHmod}

The simplest potentially realistic dependence structure in multivariate spatial models is a separable model that is a fairly simple extension of the IW-FH model described above. In the FH context of interest here, this implies considering a common spatial dependence structure that is appropriate for all outcomes, but where such outcomes share a common dependence within each spatial region. That is, we consider
$${\mbf u} \sim Gau({\mbf 0}, {\mbf \Sigma}_s \otimes {\mbf \Sigma}_o),$$
where ${\mbf \Sigma}_s$ denotes an  $n$-dimensional spatial covariance matrix and ${\mbf \Sigma}_o$ denotes an $m$-dimensional outcome covariance matrix. We refer to this as the ``Separable-FH'' model.  Even in this simplified separable framework, there are typically too many parameters to estimate an unstructured form for ${\mbf \Sigma}_s$, but for most applications, $m$ is fairly small and so it may be possible to estimate an unstructured ${\mbf \Sigma}_o$.  We thus propose a fairly simple, but potentially useful, structure in which ${\mbf \Sigma}_s$ is parameterized by the aforementioned CAR spatial dependence structure and ${\mbf \Sigma}_o$ is given an IW prior (in a Bayesian implementation). In this case, the latent random effects vector has distribution
\[\mbf{u} \sim N({\mbf 0},(\mbf{I}-\rho \mbf{W})^{-1} \otimes \mbv{\Sigma}_{IW}),\]
where ${\mbf \Sigma}_{IW}$ is given an IW prior and we would also assign a prior to $\rho$  (see below). Note, the parameter $\tau^2$ is not included in the CAR structure in this model due to identifiability issues. That is, because each element of $\mbv{\Sigma}_{IW}$ is estimated, $\tau^2$ becomes a multiplicative constant in the covariance matrix, and thus, it cannot be uniquely estimated.  In the case where all the off-diagonal elements of ${\mbf \Sigma}_{IW}$ are zero, the model reverts to the case of independent CAR models that share a common spatial dependence parameter.

Despite the seemingly limited nature of such a separable model, this model is actually quite flexible.  The IW structure allows a different conditional variance to be used for the spatial structure of each outcome, but this structure assumes that all areas have a common spatial dependency parameter, $\rho$.  This assumption would be reasonable if one had survey outcomes that are demographically related, but may not be realistic when the outcome measures are quite different. In the ICAR framework, $\rho$ is fixed at a value of 1, and in that case, there is no limitation in this aspect of the specification.  Perhaps the primary advantage of this separable formulation is that the model scales up to $m>2$ quite naturally by increasing the dimensionailty of $\mbv{\Sigma}_{IW}$ without increasing the number of parameters in the spatial dependence covariance matrix.


\subsection{The GMCAR-FH Model}\label{sec:GMCARFHmod}
Although the separable model described in Section \ref{sec:sepFHmod} can be quite useful, there may be situations for which sampling outcomes have a more complex multivariate spatial structure. In particular, we may require that the spatial structure be different for different outcomes, and that there should be more complicated cross-dependence between spatial location and outcome variable. This more complicated structure can be accommodated fairly easily in the latent effects structure. As summarized in \cite{cressiewikle2011}, there are several approaches to modeling multivariate spatial structure, but the one that has gained the most favor in recent years is the conditional approach first detailed in a hierarchical framework by \cite{royle1999hierarchical}.  The GMCAR of \cite{jin2005generalized} extends this concept of conditioning to the CAR model framework.

To reduce the notational complexity, we present the GMCAR formulation in terms of the bivariate spatial processes $\mbf{u}_1$ and $\mbf{u}_2$, as defined previously, and assume that
\[\mbf{u}_1|\mbf{u}_2 \sim N(\mbf{A} \mbf{u}_2, \tau_1^2 (\mbf{D}- \rho_1 \mbf{W})^{-1}),\]
 and
\[\mbf{u}_2  \sim N(\mbf{0}, \tau_2^2(\mbf{D}-\rho_2 \mbf{W})^{-1}),\]
where the matrix ${\mbf A}$ is parameterized in terms of two parameters, $\eta_0$ and $\eta_1$, such that $\mbf{A}=\eta_0 \mbf{I} +\eta_1 \mbf{W}$.  Thus, the conditional distribution of ${\mbf u}_1$ given ${\mbf u}_2$ is a CAR structure with mean given by ${\mbf A}{\mbf u}_2$, and the marginal distribution of ${\mbf u}_2$ is also a CAR model, but with zero mean.  Both of these distributions have unique spatial dependence parameters given by $\rho_1$ and  $\rho_2$. One can easily show that the joint distribution of ${\mbf u} = ({\mbf u}'_1, {\mbf u}'_2)'$ then follows a multivariate Gaussian process with mean ${\mbf 0}$ and covariance matrix
\[  {\mbf \Sigma}_u =  \left( \begin{array}{cc}
\tau_1^2 (\mbf{D}-\rho_1 \mbf{W})^{-1}+ \mbf{A}' \tau_2^2 (\mbf{D}-\rho_2 \mbf{W})^{-1} \mbf{A} & \mbf{A}' \tau_2^2 (\mbf{D}-\rho_2 \mbf{W})^{-1} \\
 \tau_2^2 (\mbf{D}-\rho_2 \mbf{W})^{-1} \mbf{A} & \tau_2^2 (\mbf{D}-\rho_2 \mbf{W})^{-1} \\
 \end{array} \right). \]
The advantage of this conditional approach to multivariate spatial modeling is that one is guaranteed a valid joint covariance structure so long as the conditional and marginal models are valid.  In this case, they are valid given that the CAR models are formulated as stated in the introduction of Section \ref{sec:MCARFH}.  We refer to this model as the ``GMCAR-FH'' model.

It is important to note that the GMCAR formulation directly incorporates two sources of correlation into the latent spatial effects. First, there is the direct consideration of the spatial structure in the lattice for each of $\mbf{u}_1$ and $\mbf{u}_2$ due to the CAR models. Second, the model allows correlation between the spatial random effects for each outcome through the ${\mbf A}$ matrix.  Specifically, the parameter $\eta_0$ allows for cross-correlation between the latent effects within the same areal unit, and $\eta_1$ allows for cross-correlation depending on the adjacent areas.


We note that the GMCAR structure contains many multivariate CAR models as special cases, {including the models of \cite{kim2001bivariate}, \cite{gelfand2003proper}, and two independent CAR models}, as outlined in \cite{jin2005generalized}. In particular, with $\eta_0=\eta_1=0$, the model reverts to the independent CAR model case.  This is often a case worth considering for the FH model, and it is ideal to have the flexibility for this specification, depending on the estimates one obtains from $\eta_0$ and $\eta_1$ (which, clearly would be near zero if the independent CAR models were more appropriate).  Finally, although the description of the GMCAR model presented here considers only two processes, it is possible to extend the model to $m > 2$ with added notational and computational overhead.

\section{Simulated Examples}\label{sec:sim}

To gain insight into the GMCAR-FH and Separable-FH model performance relative to the IW-FH model, we consider two simulated examples. Both simulations are designed from the perspective of the data analysis provided in Section~ \ref{sec:data}.

\subsection{Simulated Example One}
\begin{figure}
\begin{tabular}{cc}
\includegraphics[height=90mm,width=90mm,angle=-90]{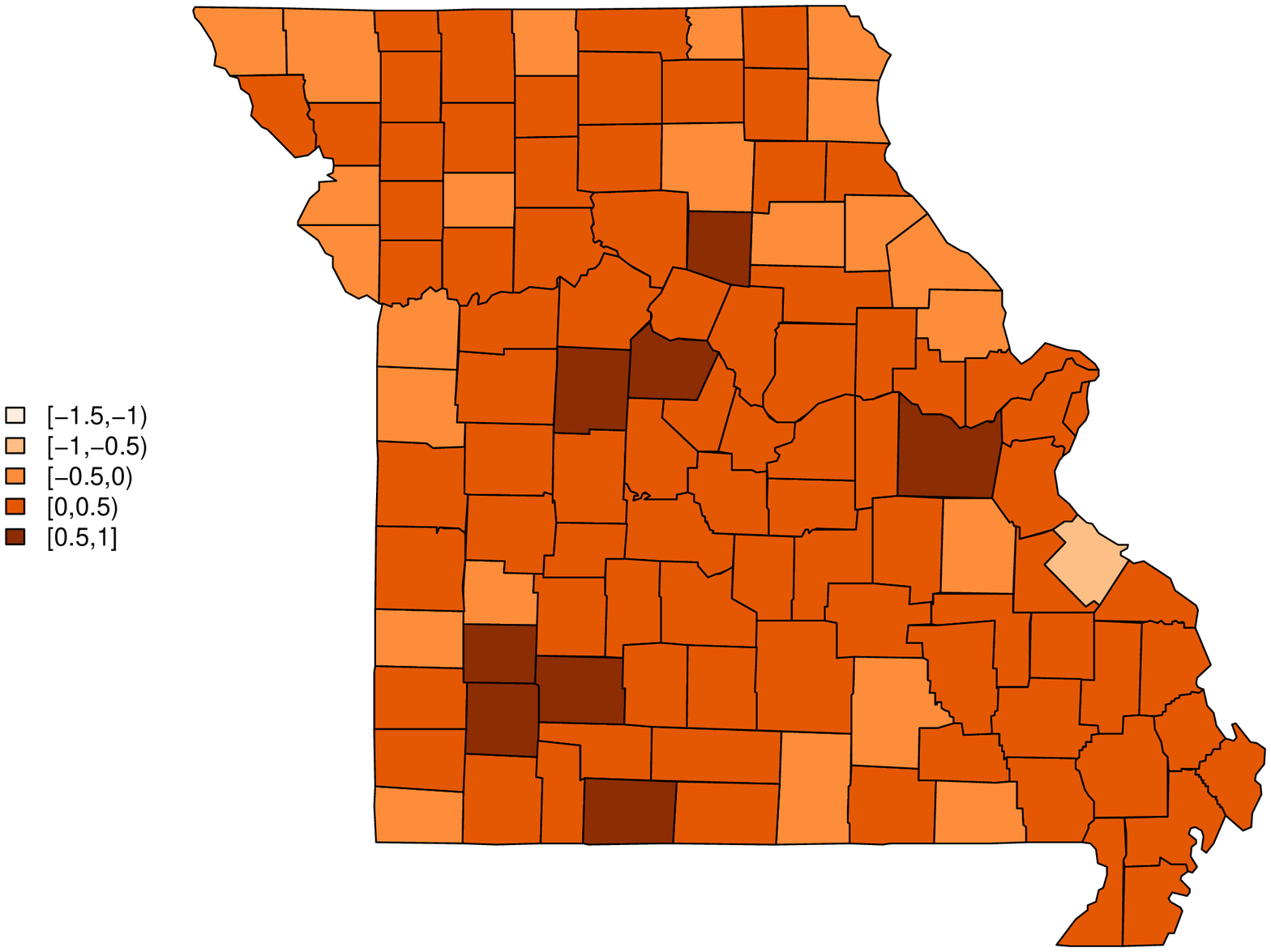} &
\includegraphics[height=90mm,width=90mm,angle=-90]{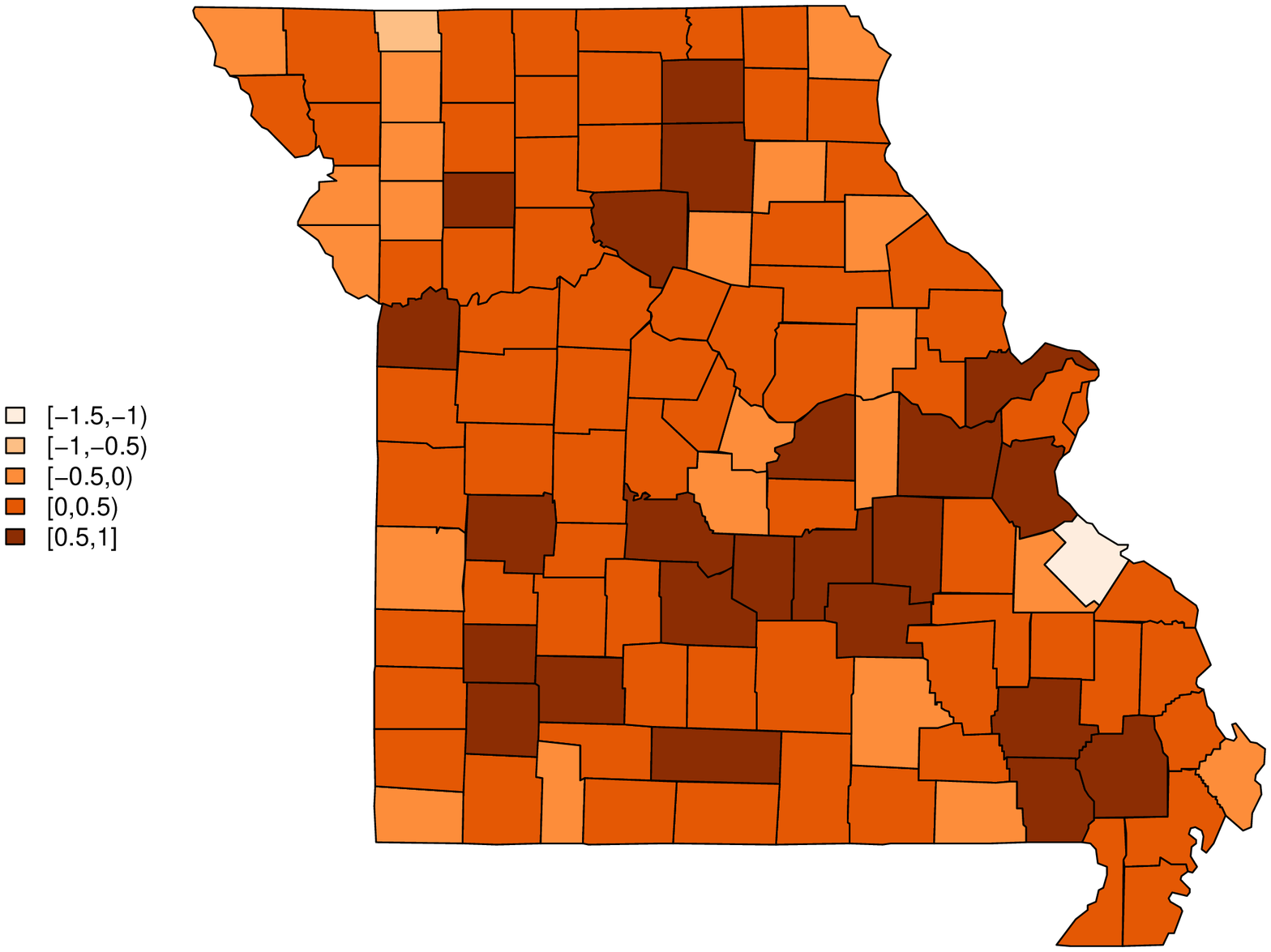}
\end{tabular}
\caption{\baselineskip=10pt Relative MSPE reduction for the GMCAR-FH model over the IW-FH model for the percentage of families in poverty (left) and percentage of individuals unemployed (right) from the simulated example in Section 3.2.}
\label{Fi:VAR}
\end{figure}

In order to assess model performance at small geographies where the true underlying values are known, we provide a simulated example using the Missouri county lattice structure (see Figure~ \ref{Fi:VAR}). We consider a bivariate outcome, generated from a GMCAR-FH structure, where the survey variances are the same as those from the data analysis in Section~ \ref{sec:data}. We utilize median household income (divided by 1000) as reported by the ACS as auxiliary information for both outcomes.

The bivariate outcome is generated according to the GMCAR-FH model of Section~\ref{sec:data}, with all parameters used for generating data set to the median values of the GMCAR-FH posterior distributions in that analysis.  Total survey variances, assumed known, are also set to be equivalent to those in the data analysis of Section~\ref{sec:data}.  Of particular interest in this analysis are the values of $\eta_0$ and $\eta_1$, which determine the cross-correlation structure. For the simulated data generation, $\eta_0$ is set to 0.26, and $\eta_1$ is set to -0.04. This results in a strong within-unit correlation and a relatively weak non-separable structure.

Model 1 uses an IW prior (see Section \ref{sec:Intro}). All the priors selected were deliberately chosen to be vague and, thus, impart little impact on our analysis. These priors are vague enough to be suitable across a wide range of analyses, and are used for all analyses presented here. We recommend $\mbox{Unif}(0,1)$ priors on the spatial parameters rather than $\mbox{Unif}(-1,1)$ priors due to the work of \cite{wall2004close}, which demonstrates that CAR models possess a variety of undesirable properties for spatial datasets with negative spatial autocorrelation. The specific prior utilized is $\mbox{IW}({\mbf I}_2,2)$, which is a vague independence prior. Model 2 is the proposed Separable-FH model, where we assume an $\mbox{IW}({\mbf I}_2,2)$ prior or ${\mbf \Sigma}_{IW}$ and a $\mbox{Unif}(0,1)$ prior for $\rho$.  Model 3 is the proposed GMCAR-FH prior, where we place relatively vague $\mbox{Unif}(0.001,100)$ priors on $\tau_1$ and $\tau_2$, $\mbox{Unif}(0,1)$ priors on $\rho_1$ and $\rho_2$, and $N(0,10^2)$ priors on $\eta_0$ and $\eta_1$.

Model comparisons are then based on posterior MSE, specifically in terms of $\mbv{\theta}_1$ and $\mbv{\theta}_2$.  These results are summarized in Table~\ref{TA:Sim1}. In this example, both the GMCAR-FH and the Separable-FH spatial models perform better than the IW-FH model in terms of MSE. The GMCAR-FH outperforms the Separable model in terms of MSE for outcome one, with both models performing similarly for outcome two. By location, the GMCAR-FH outperforms the IW-FH in terms of MSE in 70\% of locations for outcome one and 62\% of locations for outcome two, with the Separable-FH outperfoming the IW-FH model in 69\% and 65\% of locations, respectively. Additionally, the GMCAR-FH model outperforms the Separable-FH model in 52\% of locations for outcome one and the Separable-FH model outperforms the GMCAR-FH model in 51\% of locations for outcome two.

The GMCAR-FH recovers the parameters well. However, in this dataset, the true value of $\eta_1$ is -0.04, which is near enough to zero that it does not generate much non-separable structure in the model. This leads to similar performance between the GMCAR-FH and Separable-FH models in terms of MSE.

\bigskip

\begin{table}[h]
\centering
\begin{tabular}{|c|c|c|c|c|} \hline
Model & MSE1 & MSE2 \\ \hline \hline
Model 1: IW  & 0.07595 & 0.05640\\ \hline
Model 2: Separable &  0.06822& \textbf{0.05227}\\ \hline
Model 3: GMCAR & \textbf{0.06801} & 0.05230\\ \hline
\end{tabular}
\parbox{5.0in}{
\caption{The posterior MSEs for $\mbv{\theta}_1$ and $\mbv{\theta}_2$ for the IW-FH, GMCAR-FH and Separable-FH models for the Missouri county lattice simulated data described in Section 3.1. Note that MSE1 corresponds to outcome one and MSE2 corresponds to outcome two.}
\label{TA:Sim1}}
\end{table}

\subsection{Simulated Example Two}\label{sec:sim2}
In this simulated example, we utilize the exact same data generation structure as in the first simulated example, but we increase the cross correlation parameter $\eta_1$ to 0.21 in order to investigate the effects of strong cross correlation on our model. All three models were run on this new simulated data with the same prior specifications as the previous simulation.

\bigskip

\begin{table}[h]
\centering
\begin{tabular}{|c|c|c|c|c|} \hline
Model & MSE1 & MSE2 \\ \hline \hline
Model 1: IW & 0.0822 & 0.0437\\ \hline
Model 2: Separable &  0.0756 & 0.0405\\ \hline
Model 3: GMCAR &  \textbf{0.0734} & \textbf{0.0402}\\ \hline
\end{tabular}
\parbox{5.0in}{
\caption{The posterior MSEs for $\mbv{\theta}_1$ and $\mbv{\theta}_2$ for the IW-FH, GMCAR-FH and Separable-FH models for the Missouri county lattice simulated data described in Section 3.2. Note MSE1 corresponds to outcome one and MSE2 corresponds to outcome two.}
\label{TA:Sim2}}
\end{table}

The results from this simulation, summarized in Table~\ref{TA:Sim2}, clearly demonstrate an improvement in terms of MSE of the Separable-FH and GMCAR-FH models over the IW-FH model, with the GMCAR-FH providing lower MSE for the estimates of both outcomes.  The GMCAR-FH and Separable-FH models both outperform the IW-FH model in 70\% of locations in terms of MSE for both outcomes. The GMCAR-FH model additionally outperforms the Separable-FH model in 59\% of locations in terms of MSE for outcome one and 48\% of locations for outcome two. However, as is clear from Table \ref{TA:Sim2}, it performs better in terms of overall MSE.

\section{Example: State Level Estimation for Validation}\label{sec:state}

We consider the model on U.S. state-level data for which decennial census data are available. We work at the state level as these data typically exhibit spatial dependence and provide  a way to test various multivariate FH models relative to a ``known'' quantity (e.g., census data).  In particular, we predict the percentage of vacant residences and the percent of renter-occupied units in each of the contiguous 48 states and in Washington D.C. for 2010 based on American Community Survey (ACS) one-year estimates.  The 2010 decennial Census values for the same question differ in terms of residency rules (i.e., see www.census.gov).   However, we do not expect that the variables of interest here will be greatly affected by these residency rule differences and so we consider the census value as ``true" when analyzing the MSE of the IW-FH, Separable-FH, and GMCAR-FH models.  Areas with increased numbers of vacant locations are thought to be associated with economic recessions, as are areas with large percentages of renter occupied units.  Thus, we consider median household income as an auxiliary variable for both response outcomes.

In order to satisfy the normality constraint of the models considered here, the percentage of vacant properties is transformed by multiplying by 100 and then applying a log transform.  Due to the requirement in these models that the sampling errors, $\epsilon_i$, follow a normal distribution, we use the Delta method and ACS-reported margins of error (MOEs) to obtain the survey variance of the transformed variables.  We assume the errors are uncorrelated at the survey level when conditioned on the latent process as described previously.

Exploratory data analysis, in the form of linear regression, demonstrates that the auxiliary information is positively associated with the respective outcomes  ($\mbox{$p$-value} <0.05$). Further, the residuals have latent spatial structure based on a Moran's I test with a first order weight matrix ( $\mbox{$p$-value} <0.01$ for both regressions). The residuals are strongly correlated  (a test that the population correlation coefficient is non-zero yielded a $\mbox{$p$-value} =0.07$), which suggests that these are data that many researchers would find suitable for the IW-FH model. However, it is also the case that the spatial pattern in the residuals is strong, and thus, should be accounted for when modeling these data.

We perform a leave-one-out experiment where each of the 49 locations is left out of the model one at a time, and a Markov chain Monte Carlo (MCMC) algorithm is run for 15,000 iterations for each. We discard the first 1,000 iterations from each location as burn-in. Convergence of the MCMC algorithm was assessed through visual inspection of the sample chains, with no evidence of nonconvergence detected.

The mean squared error (MSE) for both outcomes at each location is then computed assuming the 2010 decennial census values to be the truth; these results are shown in Table \ref{TA:census}. In this example, the GMCAR-FH outperforms both Model 1 and Model 2, with the GMCAR-FH yielding a 35\% reduction in posterior predictive MSE for the percentage of vacant houses in the state and a 75\% reduction in posterior predictive MSE for the percentage of renter occupied residences relative to the standard IW-FH model.   As expected, the Separable-FH model performed better than the IW-FH model but not as well as the GMCAR-FH. The added spatial structure is clearly important, but the separable structure does not offer the flexibility of the GMCAR-FH model for this dataset.

\begin{table}[h]
\centering
\begin{tabular}{|c|c|c|} \hline
Model & MSE1 & MSE2 \\ \hline \hline
Model 1: IW & 0.1280 & 0.0218 \\ \hline
Model 2: Separable & 0.0974 & 0.0123 \\ \hline
Model 3: GMCAR & \textbf{0.0883} & \textbf{0.0055} \\ \hline
\end{tabular}
\parbox{5.0in}{
\caption{The posterior MSEs for the three FH Models applied to the state-level data. MSE1 is the posterior predictive MSE for the transformed percentage of vacant units, and MSE2 is the posterior predictive MSE for the transformed percentage of families below the poverty level.}}
\label{TA:census}
\end{table}

\begin{figure}
\begin{tabular}{cc}
\includegraphics[height=90mm,width=90mm,angle=-90]{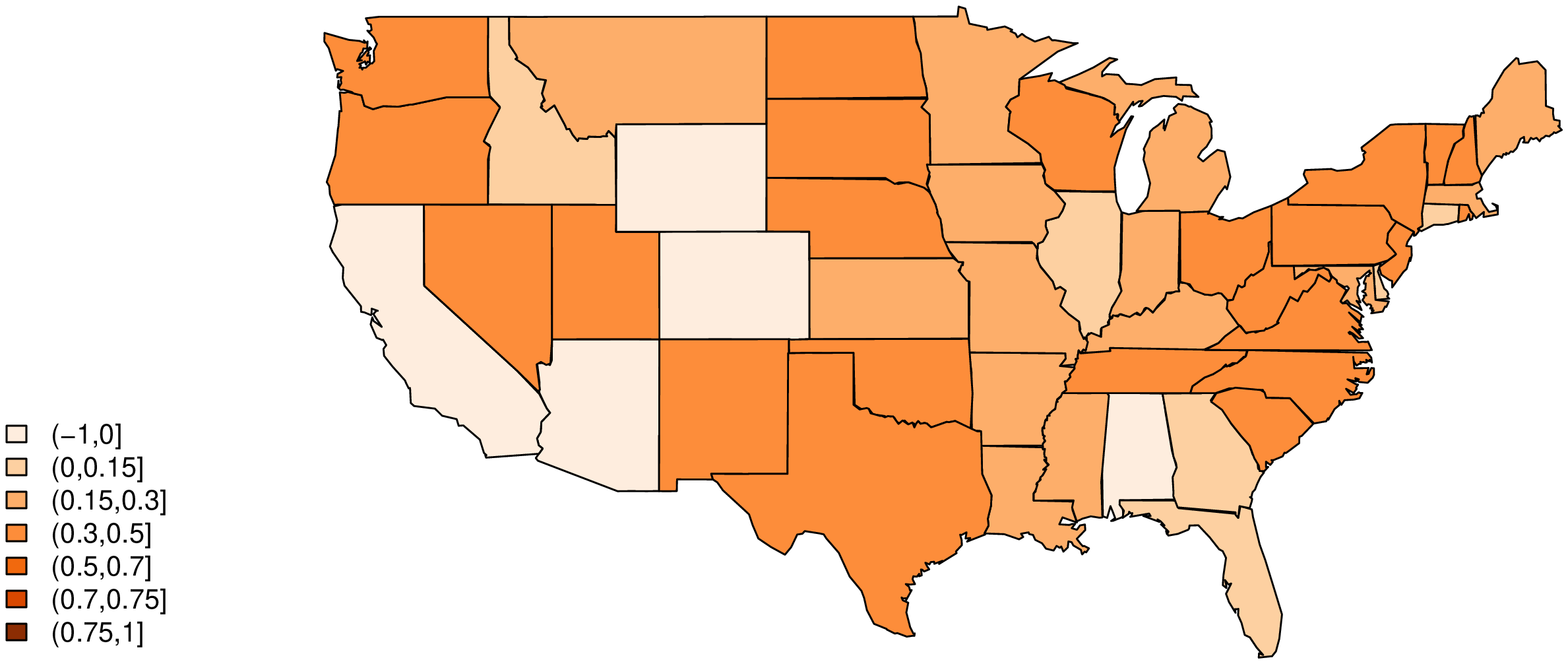} &
\includegraphics[height=90mm,width=90mm,angle=-90]{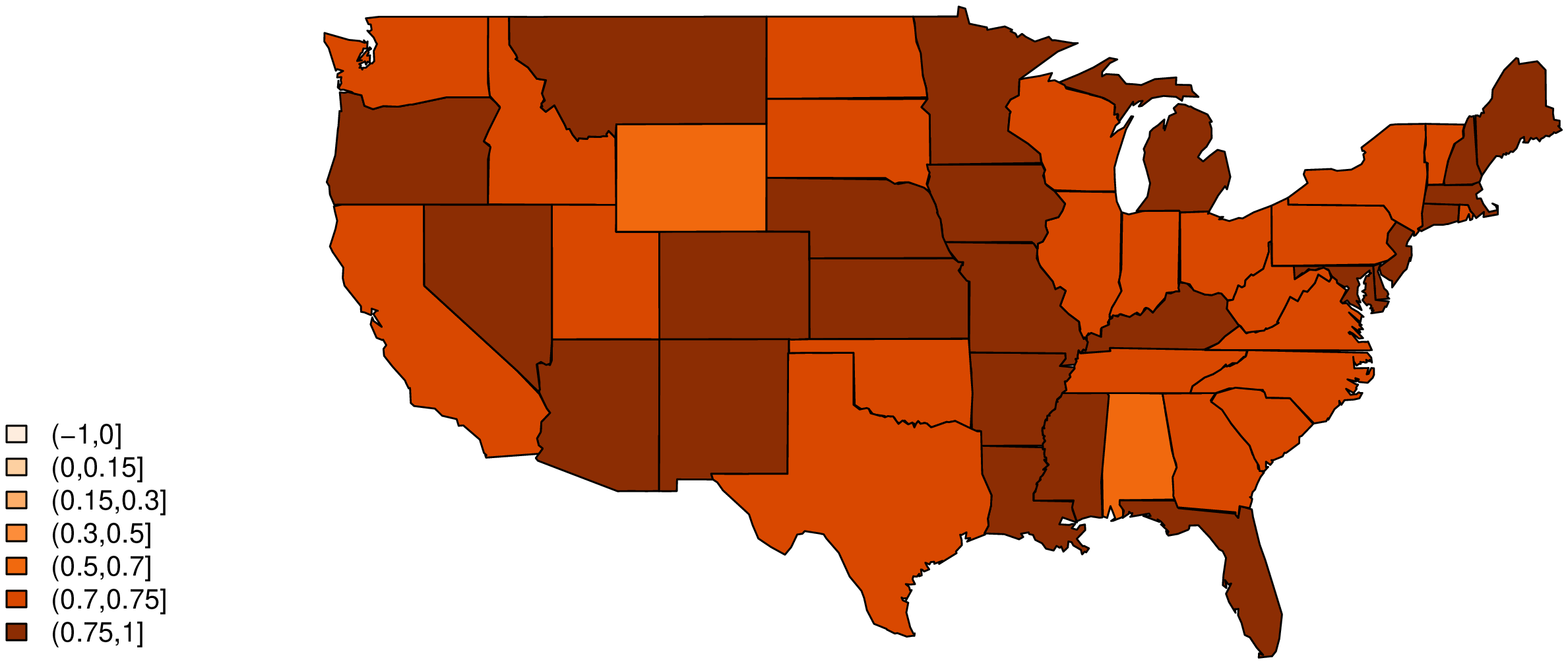}
\end{tabular}
\caption{\baselineskip=10pt MSE reduction for the GMCAR-FH model relative to the IW-FH model for the percentage of vacant units (left) and percentage renter occupied units (right) from the simulated example in Section 3.1.}
\label{Fi:MSE}
\end{figure}

Additionally, the relative improvement of the posterior predictive MSE of the GMCAR-FH model over the baseline IW-FH model is plotted as a function of location in Figure \ref{Fi:MSE}. The actual proportions are computed using the relative MSE formula given by
\begin{equation}
\nonumber \frac{\mbox{MSE}_{\mbox{\tiny{IW}}}(i)-\mbox{MSE}_{\mbox{\tiny{GMCAR}}}(i)}{\frac{1}{2}\mbox{MSE}_{\mbox{\tiny{IW}}}(i)
+\frac{1}{2}\mbox{MSE}_{\mbox{\tiny{GMCAR}}}(i)},
\end{equation}
where we define $\mbox{MSE}_{\mbox{\tiny{GMCAR}}}(i)$ to be the MSE for the GMCAR-FH model $i$ and $\mbox{MSE}_{\mbox{\tiny{IW}}}(i)$ similarly for the IW-FH model.
%
We see that the GMCAR-FH outperforms the IW-FH in all but four locations with respect to the prediction of vacant houses, and outperforms the IW-FH in every location with respect to the prediction of renter occupied residences.

\section{Example: County-Level Small Area Estimation}\label{sec:data}
To examine the model at smaller levels of geography, we consider county-level data in the state of Missouri.  We use ACS five year period estimates of the percentage of families below the poverty level as well as the percentage of unemployed individuals in each county between the years of 2006 and 2010.  Log transforms were used to obtain normally distributed outcomes, and the Delta method was used to compute the transformed survey variances, which are computed based on the MOEs reported by the ACS and assumed uncorrelated.  We consider the ACS variable median household income as auxiliary information for both outcomes.

The auxiliary information is apparently an important predictor of both outcomes based on an exploratory linear regression ($\mbox{$p$-value} <0.01$). Further, Moran's I tests with a first order weight matrix demonstrate latent spatial structure in both sets of residuals $(\mbox{$p$-value} <0.01)$ and both sets of residuals are correlated based on a Pearson correlation test $(\mbox{$p$-value} <0.05)$. Shapiro-Wilk tests did not reject the assumption of normality for both sets of residuals $(\mbox{$p$-value} >0.05)$. The three models considered in the county-level analysis were run with the same prior specifications, as given in Section \ref{sec:state}.

We compare the three FH models in terms of a leave-one-out mean squared prediction error (MSPE) experiment, where we compare our model predictions to the observed value at the left-out location.  Each model is run for 11,000 iterations, with the first 1,000 iterations discarded for burn-in. Convergence of the MCMC algorithm is assessed through visual inspection of the trace plots of the sample chains. No evidence of nonconvergence was detected. The results of the experiment can be found in Table \ref{TA:VAR}.

\begin{table}[h]
\centering
\begin{tabular}{|c|c|c|} \hline
Model & MSPE1 & MSPE2 \\ \hline \hline
Model 1: IW & 0.1367 & 0.0804 \\ \hline
Model 2: Separable  & \textbf{0.1197} & 0.0686  \\ \hline
Model 3: GMCAR &  0.1205 & \textbf{0.0684} \\ \hline
\end{tabular}
\parbox{5.0in}{
\caption{The leave-one-out MSPEs for the county level example for the three FH models. MSPE1 is the average posterior MSPE for the transformed transformed percentage of families below the poverty level, and MSPE2 is the average posterior MSPE for the percentage of unemployed individuals in each county.}
\label{TA:VAR}}
\end{table}
The Separable-FH model performs similarly to the GMCAR-FH model for this dataset. As demonstrated in the simulated examples, this is due to the near-zero value of $\eta_1$ in this dataset (i.e., posterior median $= -0.04$) and similar distributions for $\rho_1$ and $\rho_2$, which suggests that there is little nonseparable structure in this particular dataset. However, the results clearly demonstrate the importance of accounting for the multivariate spatial dependence in these data, with both models outperforming the the IW-FH for both outcomes. Figure \ref{Fi:VAR} shows the relative MSPE reduction for the GMCAR-FH model over the IW-FH model by location. We note that the GMCAR-FH model outperforms the IW-FH model in terms of MSPE in 79\% of locations for percentage of families below the poverty level and 80\% of locations for percentage of unemployed individuals, while the Separable-FH model outperforms the IW-FH in 78\% and 79\% of locations respectively.


\section{Discussion}\label{sec:disc}
In addition to MSE reduction via auxiliary covariate information, FH models can be improved by borrowing strength across spatial location as well as by the addition of multivariate relationships between outcomes. In this regard, we present two specifications for latent dependence that can be utilized to effectively incorporate the explicit CAR spatial structure within a multivariate FH framework. We have demonstrated that the proposed GMCAR-FH model shows a substantial reduction in MSE when compared to equivalent decennial census values at the state level.  In addition, for county-level SAE, we demonstrate that both the proposed GMCAR-FH and Separable-FH models show substantial reduction in MSPE over a baseline IW-FH model.  Exploratory analysis for both datasets suggested that there was correlation between both sets of residuals, which suggests that the IW-FH methodology would not be unreasonable. However, in both cases where there was additional spatial structure in the residuals, and thus, it is reasonable that our multivariate CAR structure provides improvement in estimation.

It is important to note that although the GMCAR-FH model is sometimes preferable to the Separable-FH model and the IW-FH model, it comes with increased overhead as the number of outcomes increase beyond two.  In that case, the Separable-FH model provides a more computationally convenient alternative, and the simulation results presented here suggest that it its performance with regards to MSE is typically better than the standard IW-FH model.

Because it is common that the ACS and other surveys exhibit outcomes are that spatially correlated, and it is unlikely that one will have access to all of the necessary auxiliary information to capture all of the underlying spatial structure in the residuals of the FH regressions. As such, this work provides two important classes of models for SAE in the context of federal surveys. The work extends the current thrust of explicitly accounting for spatial correlation in the FH structure to the multivariate case, which can assist in the estimation of multivariate outcomes. Future work will consider realistic spatio-temporal structures for latent effects in both the univariate and multivariate settings.

\section*{Acknowledgments}
This research was partially supported by the U.S. National Science Foundation (NSF) and the U.S. Census Bureau under NSF grant SES-1132031, funded through the NSF-Census Research Network (NCRN) program.

\clearpage\pagebreak\newpage
\bibliographystyle{apa}
\bibliography{STSN}

\begin{thebibliography}{}

\bibitem[\protect\astroncite{Besag}{1974}]{besag1974spatial}
Besag, J. (1974).
\newblock Spatial interaction and the statistical analysis of lattice systems.
\newblock {\em Journal of the Royal Statistical Society. Series B
  (Methodological)}, 36(2):192--236.

\bibitem[\protect\astroncite{Clayton and
  Bernardinelli}{1992}]{clayton1992bayesian}
Clayton, D. and Bernardinelli, L. (1992).
\newblock Bayesian methods for mapping disease risk.
\newblock {\em Geographical and environmental epidemiology: methods for small
  area studies}, pages 205--220.

\bibitem[\protect\astroncite{Clayton and Kaldor}{1987}]{clayton1987empirical}
Clayton, D. and Kaldor, J. (1987).
\newblock Empirical {B}ayes estimates of age-standardized relative risks for
  use in disease mapping.
\newblock {\em Biometrics}, pages 671--681.

\bibitem[\protect\astroncite{Cressie}{1990}]{cressie1990small}
Cressie, N. (1990).
\newblock Small area prediction of undercount using the general linear model.
\newblock In {\em Proceedings of Statistics Canada Symposium 90, Measurement
  and Improvement of Data Quality}.

\bibitem[\protect\astroncite{Cressie and Chan}{1989}]{cressie1989spatial}
Cressie, N. and Chan, N.~H. (1989).
\newblock Spatial modeling of regional variables.
\newblock {\em Journal of the American Statistical Association},
  84(406):393--401.

\bibitem[\protect\astroncite{Cressie and Wikle}{2011}]{cressiewikle2011}
Cressie, N. and Wikle, C. (2011).
\newblock {\em Statistics for Spatio-Temporal Data}.
\newblock Wiley, Hoboken, NJ.

\bibitem[\protect\astroncite{Cressie}{1993}]{cressie93}
Cressie, N. A.~C. (1993).
\newblock {\em Statistics for Spatial Data (Revised Edition)}.
\newblock Wiley-Interscience, New York.

\bibitem[\protect\astroncite{Datta et~al.}{1998}]{datta1998multivariate}
Datta, G., Day, B., and Maiti, T. (1998).
\newblock Multivariate {B}ayesian small area estimation: An application to
  survey and satellite data.
\newblock {\em Sankhya. Series A}, 60(3):344--362.

\bibitem[\protect\astroncite{Fabrizi et~al.}{2011}]{fabrizi2011hierarchical}
Fabrizi, E., Ferrante, M., Pacei, S., and Trivisano, C. (2011).
\newblock Hierarchical {B}ayes multivariate estimation of poverty rates based
  on increasing thresholds for small domains.
\newblock {\em Computational Statistics \& Data Analysis}, 55(4):1736--1747.

\bibitem[\protect\astroncite{Fay}{1987}]{fay1987application}
Fay, R. (1987).
\newblock Application of multivariate regression to small domain estimation.
\newblock {\em Small Area Statistics (R. Platek, JNK Rao, CE Sarndal and MP
  Singh, eds.)}, pages 91--102.

\bibitem[\protect\astroncite{Fay and Herriot}{1979}]{fay1979estimates}
Fay, R. and Herriot, R. (1979).
\newblock Estimates of income for small places: an application of
  {J}ames-{S}tein procedures to census data.
\newblock {\em Journal of the American Statistical Association}, 74:269--277.

\bibitem[\protect\astroncite{Gelfand and Vounatsou}{2003}]{gelfand2003proper}
Gelfand, A.~E. and Vounatsou, P. (2003).
\newblock Proper multivariate conditional autoregressive models for spatial
  data analysis.
\newblock {\em Biostatistics}, 4(1):11--15.

\bibitem[\protect\astroncite{Gomez-Rubio et~al.}{2010}]{gomez2010bayesian}
Gomez-Rubio, V., Best, N., Richardson, S., Li, G., and Clarke, P. (2010).
\newblock {Bayesian Statistics Small Area Estimation}.
\newblock Technical report, Imperial College London.

\bibitem[\protect\astroncite{Gonz{\'a}lez-Manteiga
  et~al.}{2008}]{gonzalez2008analytic}
Gonz{\'a}lez-Manteiga, W., Lombard{\'\i}a, M., Molina, I., Morales, D., and
  Santamar{\'\i}a, L. (2008).
\newblock Analytic and bootstrap approximations of prediction errors under a
  multivariate fay--herriot model.
\newblock {\em Computational Statistics \& Data Analysis}, 52(12):5242--5252.

\bibitem[\protect\astroncite{Huang and Bell}{2004}]{huang2004empirical}
Huang, E. and Bell, W. (2004).
\newblock An empirical study on using acs supplementary survey data in saipe
  state poverty models.
\newblock In {\em 2004 Proceedings of the American Statistical Association},
  pages 3677--3684.

\bibitem[\protect\astroncite{Jin et~al.}{2005}]{jin2005generalized}
Jin, X., Carlin, B., and Banerjee, S. (2005).
\newblock Generalized hierarchical multivariate {C}{A}{R} models for areal
  data.
\newblock {\em Biometrics}, 61(4):950--961.

\bibitem[\protect\astroncite{Kim et~al.}{2001}]{kim2001bivariate}
Kim, H., Sun, D., and Tsutakawa, R. (2001).
\newblock A bivariate {B}ayes method for improving the estimates of mortality
  rates with a twofold conditional autoregressive model.
\newblock {\em Journal of the American Statistical Association}, 96:1506--1521.

\bibitem[\protect\astroncite{Kubokawa and
  Nagashima}{2011}]{kubokawa2011parametric}
Kubokawa, T. and Nagashima, B. (2011).
\newblock Parametric bootstrap methods for bias correction in linear mixed
  models.
\newblock {\em Journal of Multivariate Analysis}.

\bibitem[\protect\astroncite{Leroux et~al.}{1999}]{leroux1999estimation}
Leroux, B., Lei, X., and Breslow, N. (1999).
\newblock {Estimation of Disease Rates in Small Areas: A New Mixed Model for
  Spatial Dependence}.
\newblock In {\em Statistical Models in Epidemiology, the Environment and
  Clinical Trials}, volume 116, pages 135--178. Springer.

\bibitem[\protect\astroncite{MacNab}{2003}]{macnab2003hierarchical}
MacNab, Y. (2003).
\newblock Hierarchical {B}ayesian spatial modelling of small-area rates of
  non-rare disease.
\newblock {\em Statistics in Medicine}, 22(10):1761--1773.

\bibitem[\protect\astroncite{Mardia}{1988}]{mardia1988multi}
Mardia, K. (1988).
\newblock Multi-dimensional multivariate {G}aussian {M}arkov random fields with
  application to image processing.
\newblock {\em Journal of Multivariate Analysis}, 24(2):265--284.

\bibitem[\protect\astroncite{Mollie and Richardson}{1991}]{mollie1991empirical}
Mollie, A. and Richardson, S. (1991).
\newblock Empirical bayes estimates of cancer mortality rates using spatial
  models.
\newblock {\em Statistics in Medicine}, 10(1):95--112.

\bibitem[\protect\astroncite{Pratesi and Salvati}{2008}]{pratesi2008small}
Pratesi, M. and Salvati, N. (2008).
\newblock Small area estimation: the {E}{B}{L}{U}{P} estimator based on
  spatially correlated random area effects.
\newblock {\em Statistical Methods and Applications}, 17(1):113--141.

\bibitem[\protect\astroncite{Rao}{2003}]{rao2003small}
Rao, J. (2003).
\newblock {\em Small {A}rea {E}stimation}.
\newblock Wiley-Interscience, Hoboken, NJ.

\bibitem[\protect\astroncite{Royle and Berliner}{1999}]{royle1999hierarchical}
Royle, J.~A. and Berliner, L.~M. (1999).
\newblock A hierarchical approach to multivariate spatial modeling and
  prediction.
\newblock {\em Journal of Agricultural, Biological, and Environmental
  Statistics}, 4:29--56.

\bibitem[\protect\astroncite{Rue and Held}{2005}]{rue2005gaussian}
Rue, H. and Held, L. (2005).
\newblock {\em Gaussian {M}arkov {R}andom {F}ields: {T}heory and
  {A}pplications}.
\newblock Chapman \& Hall/CRC, Boca Raton, FL.

\bibitem[\protect\astroncite{Sain et~al.}{2011}]{sain2011spatial}
Sain, S.~R., Furrer, R., and Cressie, N. (2011).
\newblock A spatial analysis of multivariate output from regional climate
  models.
\newblock {\em The Annals of Applied Statistics}, 5(1):150--175.

\bibitem[\protect\astroncite{Torabi}{2011}]{torabi2011hierarchical}
Torabi, M. (2011).
\newblock Hierarchical {B}ayes estimation of spatial statistics for rates.
\newblock {\em Journal of Statistical Planning and Inference},
  142(1):358--–365.

\bibitem[\protect\astroncite{Wall}{2004}]{wall2004close}
Wall, M. (2004).
\newblock A close look at the spatial structure implied by the {C}{A}{R} and
  {S}{A}{R} models.
\newblock {\em Journal of Statistical Planning and Inference}, 121(2):311--324.

\bibitem[\protect\astroncite{You and Rao}{2002}]{you2002small}
You, Y. and Rao, J. (2002).
\newblock Small area estimation using unmatched sampling and linking models.
\newblock {\em Canadian Journal of Statistics}, 30(1):3--15.

\bibitem[\protect\astroncite{You and Zhou}{2011}]{you2011hierarchical}
You, Y. and Zhou, Q. (2011).
\newblock Hierarchical {B}ayes small area estimation under a spatial model with
  application to health survey data.
\newblock {\em Survey Methodology}, 37(1):25--36.

\end{thebibliography}

\clearpage\pagebreak\newpage

\baselineskip=14pt \vskip 4mm\noindent

\baselineskip=14pt \vskip 4mm\noindent

\end{document}